\documentclass[aps,pre,reprint,twocolumn, footinbib,superscriptaddress]{revtex4-1}

\usepackage{amssymb,amsmath}
\usepackage{graphicx}
\usepackage{color}
\usepackage{listings}
\usepackage{float}
\usepackage{wrapfig}
\usepackage{physics}
\usepackage{amsmath}
\usepackage{gensymb}
\usepackage{soul}
\usepackage{caption}
\usepackage{subcaption}

\usepackage{upgreek}
\usepackage{enumerate} 
\usepackage[linkcolor = blue, citecolor = blue, urlcolor = blue, colorlinks = true]{hyperref}

\usepackage[version=3]{mhchem}
\usepackage{commath,amssymb}
\usepackage{ushort}
\usepackage{multirow}
\usepackage[usenames,dvipsnames]{xcolor}
\usepackage{cleveref}
\usepackage{epstopdf}
\usepackage[exponent-product = \cdot,separate-uncertainty]{siunitx}

\crefname{equation}{Eq.}{Eqs.}
\crefname{figure}{Fig.}{Figs.}
\crefrangeformat{equation}{Eqs.~#3(#1)#4--#5(#2)#6}

\colorlet{darkgreen}{green!40!black}

\hypersetup{
    colorlinks=true,
    linkcolor=blue,
    citecolor=red}

\begin{document}

\title{\textbf{Height Distribution and Orientation of Colloidal Dumbbells Near a Wall}}

\author{Ruben W. Verweij} 
\altaffiliation{These authors contributed equally to this work}
\author{Stefania Ketzetzi} 
\altaffiliation{These authors contributed equally to this work}
\affiliation{\small Huygens-Kamerlingh Onnes Laboratory, Leiden University, P.O. Box 9504, 2300 RA Leiden, The Netherlands}
\author{Joost de Graaf}
\affiliation{\small Institute for Theoretical Physics, Center for Extreme Matter and Emergent Phenomena, Utrecht University, Princetonplein 5, 3584 CC Utrecht, The Netherlands}
\author{Daniela J. Kraft} 
\email[Corresponding email: ]{kraft@physics.leidenuniv.nl}

\affiliation{\small Huygens-Kamerlingh Onnes Laboratory, Leiden University, P.O. Box 9504, 2300 RA Leiden, The Netherlands}

\date{\today}

\begin{abstract}
Geometric confinement strongly influences the behavior of microparticles in liquid environments. However, to date, nonspherical particle behaviors close to confining boundaries, even as simple as planar walls, remain largely unexplored. Here, we measure the height distribution and orientation of colloidal dumbbells above walls by means of digital in-line holographic microscopy. We find that while larger dumbbells are oriented almost parallel to the wall, smaller dumbbells of the same material are surprisingly oriented at preferred angles. We determine the total height-dependent force acting on the dumbbells by considering gravitational effects and electrostatic particle-wall interactions. Our modeling reveals that at specific heights both net forces and torques on the dumbbells are simultaneously below the thermal force and energy, respectively, which makes the observed orientations possible. Our results highlight the rich near-wall dynamics of nonspherical particles, and can further contribute to the development of quantitative frameworks for arbitrarily-shaped microparticle dynamics in confinement.
\end{abstract}

\maketitle

\section{Introduction}
The behavior of micron-sized colloidal particles under confinement has been a subject of intensive research in engineering, materials science, and soft matter physics~\cite{Lowen2001}. Such particles often serve as model systems for understanding the effects of confinement on microscale processes, \textit{e.g.} structure formation and rheology, offering quantitative insights into the behavior of biological systems~\cite{Kim2010,Wu2017,Han2019}. This understanding is further desirable for various applications where confinement dictates the dynamics, ranging from improving microfluid transport in lab-on-a-chip devices~\cite{Ozturk2015}, growing low-defect photonic crystals~\cite{Miguez2003}, and tuning pattern formation for materials design~\cite{Serna2020, Yang2001,Mondal2020}.

Confinement can strongly affect hydrodynamic and electrostatic (self-)interactions. These effects depend on particle-wall separation as well as particle size and shape~\cite{Wu2005}. Yet, the majority of research has focused  on the behavior of spherical particles, both from a theoretical and experimental standpoint. This includes the behavior of single spheres close to a planar wall~\cite{Lorentz1907,Faxen1922,Faxen1924,Brenner1961,Goldman1967,Frej1993,Sharma2010,Rogers2012, Huang2015}, between two walls~\cite{Lobry1996,Lin2000,Dufresne2001, Benesch2003} and microchannels~\cite{Zembrzycki2012,Dettmer2014}. Going beyond single particle dynamics, the collective behavior of sphere clusters and dense suspensions has also been examined close to~\cite{Lele2011,Michailidou2009}, as well as in between walls~\cite{Pesche2000}, microchannels~\cite{Eral2010, Cui2002} and confining droplets~\cite{Wang2019clusters}.

However, microparticles involved in biological processes and industrial applications typically depart from the ideal spherical shape. Since the motion of nonspherical particles is different from that of spherical ones~\cite{Happel1983,Han2006,Padding2010,Kraft2013, Larodo2019}, there is a need to study the effect of confinement on nonspherical particles~\cite{Haghighi2013} to gain proper understanding of both naturally occurring and technologically relevant systems. For nonspherical colloids, dynamics have typically been measured far from walls~\cite{Kraft2013}. Despite predictions for axisymmetric particles~\cite{Lisicki2016} and simulated studies for arbitrary shapes~\cite{Delong2015,Fernandes2002}, the effect of particle-wall separation remains experimentally unexplored. Yet, the interplay between shape anisotropy and wall separation ought to be examined as well, to develop accurate model systems for molecular matter.

To date, a plethora of techniques has been employed for colloidal studies, including optical microscopy~\cite{Tinoco2007}, optical tweezers~\cite{Lin2000, Leach2009, Jeney2008, Schaffer2007}, light scattering~\cite{Garnier1991, Holmqvist2007, Watarai2014, Feitosa1991}, evanescent wave dynamic light scattering~(EWDLS)~\cite{Lan1986, Holmqvist2006, Lobry1996, Michailidou2009, Lisicki2014,Kazoe2011}, total internal reflection microscopy~(TIRM)~\cite{Frej1993, Prieve1992, Volpe2009}, TIRM combined with optical tweezers~\cite{Liu2014}, holographic microscopy~\cite{Lee2007, Dixon2011}, and holographic optical tweezers~\cite{Lele2011}. Each of these techniques has its own strengths and weaknesses, especially when it comes to measuring anisotropic particle dynamics near walls with high spatiotemporal resolution in three dimensions. For example, optical microscopy is a straightforward technique, yet lacks sensitivity to out-of-plane motion. Confocal microscopy on the other hand provides accurate three-dimensional measurements, but is relatively slow when recording image stacks and additionally requires refractive index matching and fluorescent labelling. Optical tweezers confine particle motion and hence hinder long-term three-dimensional measurements, while light scattering determines ensemble properties and is thus difficult to interpret in the case of anisotropic particles~\cite{Bolintineanu2014}. TIRM is an elaborate technique that provides high resolution, though its range is limited to the near-wall regime, typically less than \SI{400}{\nm} from the wall~\cite{Frej1993, Prieve1992, Volpe2009, Liu2014}. 

To overcome the above limitations, holographic microscopy may be employed instead, as it records both position and shape~\cite{Middleton2019} with high resolution~\cite{Dixon2011}, also in the out-of-plane direction. In addition, it is even capable of resolving weakly-scattering objects as used in biology~\cite{Garcia2006,Marquet2005, Lee2007, Guiliano2014} without the need for fluorescent labeling~\cite{Xu2001}. Moreover, while measurements are typically performed using lasers, a cost-effective holographic microscopy setup can also be constructed using an LED mounted on an existing microscope~\cite{Guiliano2014}. As a downside, analyzing holographic measurements may be computationally expensive which, if desired, can be compensated by implementation of a neural network~\cite{Altman2020} at the expense of some accuracy loss. 

In this article, we measure colloidal dumbbell dynamics above a planar wall, a simple model system that enables the study of the effects of shape anisotropy on confined dynamics. We accurately probe how the particle orientation is affected by the presence of the wall, and specifically, the particle-wall separation by means of digital in-line holographic microscopy. We find that smaller dumbbells are oriented at nonzero angles with respect to the wall, while in contrast, larger dumbbells of the same material are oriented mostly parallel to the wall. In all cases, we were able to identify the relation between particle orientation and particle-wall separation. We further compare our experimental findings to a minimal model for the dumbbell that combines gravitational and electrostatic dumbbell-wall interactions. We find that, despite its simplicity, the model provides qualitative insight into our observations. Our results highlight the importance of wall effects on anisotropic particle motion, and may ultimately contribute to the development of a quantitative framework for the dynamics of particles with arbitrary shapes in confinement, not fully established at present in the literature.

\section{Methods}

\begin{figure*}
    \centering
    \includegraphics{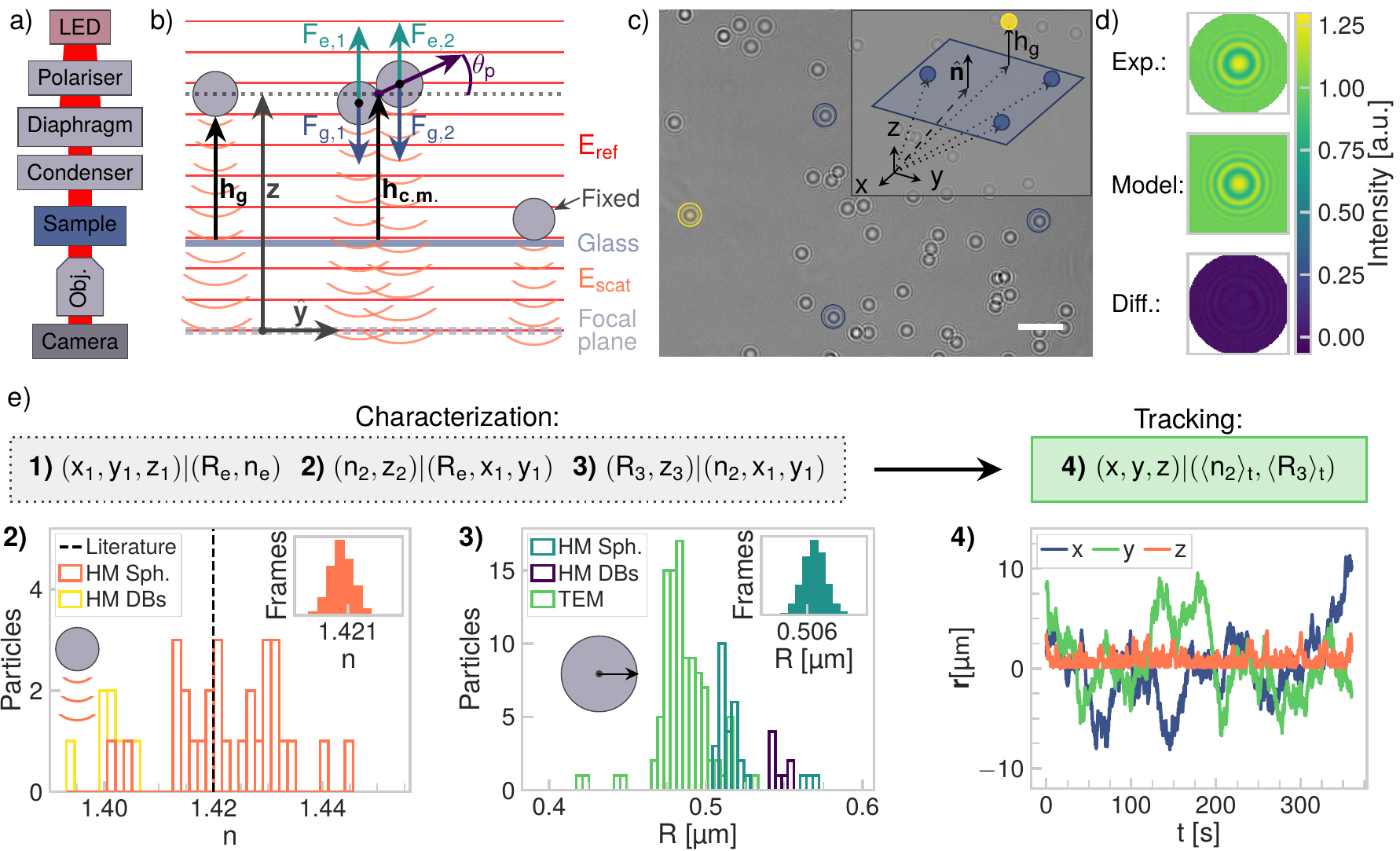}
    \caption{\textbf{Measuring particle-wall separation with in-line holographic microscopy (HM).} \textbf{a)} Schematic representation of the light path of our setup. \textbf{b)} Holograms are formed by the interference of the reference field $\mathrm{E_{ref}}$ with the scattered field $\mathrm{E_{scat}}$. We are interested in the gap height \textit{h$_g$} (or equivalently the center of mass (c.m.) height \textit{h$_{c.m.}$}) with respect to a planar glass wall. \textbf{c)} We determine the position of the wall by fitting a plane to the positions of at least three particles fixed on the wall (blue). The gap height \textit{h$_g$} between a diffusing particle (yellow) and the wall is the distance between the particle's measured position and its position projected on the plane along $\mathrm{\hat{n}}$. \textbf{d)} Comparison of an experimental image, the fitted model and the residual for a sphere, the low values of which indicate the good agreement between experimental data and model. \textbf{e)} The 3D position of the particles in time is fitted in four steps: the first three are characterization steps, in which we find the approximate 3D position (fitting step 1) as well as appropriate guesses for the refractive index $n$ (fitting step 2) and radius $R$ (fitting step 3). In the fourth step, we use these positions and the average $n$ and $R$ values to determine the 3D position accurately (fitting step 4). All steps are explained in detail in \autoref{Analysis:Spheres}. \textbf{Plot e 2)} Average $n$ obtained from fitting step 2 for both spheres and dumbbells, the inset shows a distribution from a single measurement. \textbf{Plot e 3)} Average $R$ obtained from fitting step 3 for both spheres and dumbbells, the inset shows a distribution from a single measurement. For comparison, we show particle radii measured using TEM. \textbf{Plot e 4)} Final 3D position in time for an $R=$\SI{0.55}{\um} sphere, as obtained in fitting step 4.\label{fig:fig1}}
\end{figure*}

\subsection{Materials}

We used spherical silica particles of diameter \SI{1.1\pm0.04}{\um} (size polydispersity (PD) \SI{3.7}{\%})~\footnote{The \SI{1.1}{\um} spherical silica particles and the TEM images used here were provided by Dr. Samia Ouhajji} prepared following the method of Ref.~\cite{Zhang2003silica}. Briefly, \SI{0.5}{\milli\liter} te\-tra\-e\-thyl or\-tho\-si\-li\-cate (TEOS) diluted with \SI{2}{\milli\liter} ethanol was added to a mixture of \SI{50}{\milli\liter} ethanol and \SI{10}{\milli\liter} ammonia (\SI{25}{\percent}). The mixture was stirred magnetically for \SI{2}{\hour}. The seed particles were grown to the desired size by adding \SI{5}{\milli\liter} TEOS diluted with \SI{20}{\milli\liter} ethanol during \SI{2}{\hour} using a peristaltic pump. The dispersion was stirred overnight and washed by centrifuging and redispersing in ethanol three times. We obtained their diameter and PD from transmission electron micrographs using ImageJ~\cite{Rasband1994ImageJ}, by fitting particle diameters with the software's built-in functions.

In addition, we used \SI{2.1\pm0.06}{\um} diameter (PD \SI{2.8}{\%}) spherical silica particles purchased from Microparticles GmbH. In all experiments, dumbbell particles are naturally occurring aggregates of two spherical particles. All solutions were prepared with fresh ultra-pure Milli-Q water (Milli-Q Gradient A10, \SI{18.2}{\mega\ohm\centi\meter} resistivity). Glass cover slips were purchased from VWR and were used as received.

\subsection{Holographic Setup}

We employed a digital inline holographic microscopy (DIHM) setup based on existing examples~\cite{Guiliano2014}. Our setup made use of an inverted microscope (Nikon Ti-E) equipped with a \SI{60}{\times} oil immersion objective ($\mathrm{NA} = 1.4$). To generate a scatter pattern, we used a \SI{660}{\nm} light-emitting diode (LED) source (Thorlabs M660L4) at its maximum power (\SI{3120}{\milli\watt}, using a Thorlabs LEDD1B LED driver), mounted on the lamphouse port of the microscope instead of the standard bright-field lamp (see~\autoref{fig:fig1}a for a schematic). Prior to each measurement, we performed a K\"{o}hler illumination procedure in bright-field mode to align the diaphragm and condenser. Additionally, we employed a linear polarizer on top of the condenser to improve the quality of the holograms by enforcing a specific polarization direction.

\subsection{Sample Preparation and Measurement Details}

Spherical silica particles of either 1.1 or \SI{2.1}{\um} diameter were spin coated from ethanol at dilute concentration onto the glass cover slips, which fixated their position. The cover slips were then placed at the base of the sample holder, serving as the walls relative to which particle motion was measured. The fixated-to-the-wall spheres served as reference points for determining the position of said wall (see also \autoref{fig:fig1}b and \ref{fig:fig1}c as well as the discussion in \autoref{Analysis:Plane}). Afterwards, an aqueous dispersion of particles of the same size was added in the sample holder, which was subsequently entirely filled with water and covered at the top with a glass cover slip to prevent drift. The dispersion contained single spheres as well as small fractions of dumbbell particles that consisted of two touching spheres, see also~\autoref{fig:fig1}b for an illustration. The motion of all particles above the wall was recorded at a frame rate of 19 fps for at least 6 minutes. 

\subsection{Analysis of holograms}

For all measurements, the recorded holographic microscopy images were corrected with background as well as dark-field images to minimize errors stemming from interfering impurities along the optical train. Then, for each measurement, the particle of interest was selected manually and a circular crop around its hologram was taken, see also~\autoref{fig:fig1}d, to reduce the amount of pixels considered during model fitting, thereby increasing computational efficiency. From the holograms, we determine the three-dimensional position, (\textit{x}, \textit{y}, \textit{z}), the radius, \textit{R}, and refractive index, \textit{n}, of the spheres and dumbbells as described in subsections~\ref{Analysis:Spheres} and~\ref{Analysis:Dumbbells}, respectively. 

\subsubsection{Spherical Particles}\label{Analysis:Spheres}

To fit the experimental data, we performed least-squares fits of a model based on Mie scattering theory~\cite{Lee2007} using the Python package HoloPy~\cite{barkley2018holographic} (see \autoref{fig:fig1}d as an example).

The 3D position of the particles in time was fitted in four steps (depicted in \autoref{fig:fig1}e): the first three are characterization steps to find the approximate 3D position (fitting step 1) as well as appropriate guesses for the refractive index $n$ (fitting step 2) and the radius $R$ (fitting step 3). In the fourth step, we used these positions and the average values of the radius and refractive index to determine the 3D position accurately (fitting step 4). We will now discuss these steps in detail. The subscripts correspond to the fitting step in which each parameter was determined.
\paragraph*{Fitting step 1)} For each frame, we determined the rough particle position $(x_1, y_1, z_1)$, using reasonable estimates for the radius $R_e$ and refractive index $n_e$.
\paragraph*{Fitting step 2)} For the current frame, we determined $z_2$ and characterized the particle refractive index $n_2$, while keeping the $(x_1,y_1)$ position and the estimated radius $R_e$ fixed. Example distributions and average values of the refractive indices obtained in this fitting step are shown in \autoref{fig:fig1} Plot e 2.
\paragraph*{Fitting step 3)} Whilst keeping the $(x_1,y_1)$ position and the estimated refractive index $n_2$ fixed, we fitted $z_3$ and the radius $R_3$. Example distributions and average values of the radii obtained in this fitting step are shown in \autoref{fig:fig1} Plot e 3.
\paragraph*{Fitting step 4)} Once the initial positions (\textit{x$_1$}, \textit{y$_1$}, \textit{z$_3$}) and particle properties (\textit{n$_2$}, \textit{R$_3$}) were determined for all frames, we calculated the time averaged over all frames properties ($\langle$\textit{n$_2$}$\rangle_t$, $\langle$\textit{R$_3$}$\rangle_t$). Lastly, we performed a least-squares fit for each frame allowing (\textit{x}, \textit{y}, \textit{z}) to vary, keeping ($\textit{n} = \langle\textit{n$_2$}\rangle, \textit{R} = \langle\textit{R$_3$}\rangle$) fixed (\autoref{fig:fig1} Plot e 4).

Following this procedure, we minimized unwanted correlations between (\textit{z}, \textit{R}, \textit{n}) that can arise when allowing all parameters to vary at once during the fit. For every frame, save the initial one, we used the values of the previous frame as starting guesses to speed up the (convergence of the) analysis.

\subsubsection{Dumbbell Particles}\label{Analysis:Dumbbells}

The steps followed to obtain particle properties and positions of the dumbbells were analogous to those of the single spheres, only modified to additionally account for determining the dumbbell orientations. The scattering pattern of the dumbbell, calculated using the T matrix (or null-field) method~\cite{Mackowski1996}, was modelled using the Python package HoloPy~\cite{barkley2018holographic}. We used three characterization fitting steps to find the approximate 3D position and orientation (fitting step 1) as well as appropriate guesses for refractive indices $n^{(A)}, n^{(B)}$ (fitting step 2) and the radii $R^{(A)}, R^{(B)}$ (fitting step 3). $R^{(A)}, R^{(B)}$ are the radii of the respective `A' and `B' spheres of the dumbbell with refractive indices $n^{(A)}, n^{(B)}$. In the fourth and final step, we used these positions, orientations and the average values of the radii and refractive indices to determine the 3D position and orientation accurately (fitting step 4). We will now discuss these steps in detail. The subscripts correspond to the fitting step in which each parameter was determined.
\paragraph*{Fitting step 1)} In this first step, we determined $(x_1, y_1, z_1, \alpha_1, \beta_1, \gamma_1)$ of the center-of-mass (c.m.), with $(R_e^{(A)}, n_e^{(A)}, R_e^{(B)}, n_e^{(B)})$ set to reasonable estimates. Here, $(\alpha, \beta, \gamma)$ correspond to the three Euler angles using the ZYZ convention, while $(x, y, z)$ denote the c.m. positions and, again, numbered subscripts the fitting step in which the parameter was obtained.
\paragraph*{Fitting step 2)} We determined the refractive indices and $z$-position $(n_2^{(A)}, n_2^{(B)}, z_2)$ while keeping $(x_1, y_1, \alpha_1, \beta_1, \gamma_1, R_e^{(A)}, R_e^{(B)})$ fixed.
\paragraph*{Fitting step 3)} Radii and $z$-position $(R_3^{(A)}, R_3^{(B)}, z_3)$ were fitted while $(x_1, y_1, \alpha_1, \beta_1, \gamma_1, n_2^{(A)}, n_2^{(B)})$ were kept constant.
\paragraph*{Fitting step 4)} After determining the initial positions $(x_1, y_1, z_3)$, orientations $(\alpha_1, \beta_1, \gamma_1)$ and particle properties $(n_2^{(A)}, n_2^{(B)}, R_3^{(A)}, R_3^{(B)})$ for all frames, we calculated the time averaged properties $(n^{(A)} = \langle n_2^{(A)} \rangle_t, n^{(B)} = \langle n_2^{(B)} \rangle_t, R^{(A)} = \langle R_3^{(A)} \rangle_t, R^{(B)} = \langle R_3^{(B)} \rangle_t)$ over all frames. Then, we performed a least-squares fit for each frame again, where we allowed $(x, y, z, \alpha, \beta, \gamma)$ to vary, keeping $(R^{(A)}, R^{(B)}, n^{(A)}, n^{(B)})$ fixed.

Following this procedure, we minimize unwanted correlations between $(\alpha, \beta, \gamma, z, R^{(A)}, R^{(B)}, n^{(A)}, n^{(B)})$ that can arise when allowing all parameters to vary at the same time. For every frame, save the initial one, we used the values of the previous frame as starting guesses to speed up the analysis. On that note, we additionally restricted the differences in rotation angles between subsequent frames to be smaller than \SI{90}{deg}. Finally, we used the open-source TrackPy implementation \cite{trackpy} of the Crocker-Grier algorithm \cite{Crocker1996} to link the individual sphere positions between frames into continuous trajectories, ensuring a correct and consistent orientation of the dumbbell. Because we assign specific labels to both particles in the first frame of the video, we can distinguish the particles, and in turn, between positive and negative orientations, throughout the video.

\subsection{Particle-Plane Separation}\label{Analysis:Plane}

The position and orientation of the wall was accurately determined from the three-dimensional positions of at least three spin coated spheres that were irreversibly fixed to the wall. This served two purposes: first, to speed up the fit of the mobile particles under study by providing a reliable lower bound on their axial position, and second, to accurately determine their height from the wall. A reference point on the plane $\mathbf{r}_p = (0, 0, z_p)$ and a normal vector $\mathbf{\hat{n}}_p$ (see the inset of \autoref{fig:fig1}c) were determined for all the fixed particles for each frame. Using $\mathbf{r}_p$ and $\mathbf{\hat{n}}_p$, the particle-plane separation along the normal vector $\mathbf{\hat{n}}_p$ was determined for the mobile spheres (see also~\autoref{fig:fig2}a) from~$\mathbf{\hat{n}}_p \cdot (\mathbf{r} - \mathbf{r}_p) - R$, with $\mathbf{r}$ and $R$ the position and radius of the sphere, respectively. For the dumbbells, particle-plane separation was determined using the same procedure as the individual spheres; both the c.m. height, $h_{c.m.} = \mathbf{\hat{n}}_p \cdot (\mathbf{r}_{c.m.} - \mathbf{r}_p)$, above the wall is reported, as well as gap heights of both the lower and upper sphere. Note that since the orientation of the dumbbell can flip, the lower (or upper) sphere is not necessarily always the same physical particle.

\subsection{Sphere Height Distribution}\label{Analysis:HeightDistribution}

To model the height distributions of the spherical particles above the wall in~\autoref{Sec:spheres}, we used a model that combines electrostatic and gravitational effects~\cite{Wu2005, flicker1993measuring} to calculate the total height-dependent force $F(h_{c.m.})$ in the z direction (see also schematic in~\autoref{fig:fig1}b):
\begin{align}
    &F(h_{c.m.}) = F_{e}(h_{c.m.}) + F_{g} \label{eq:model_net} \\
    &F_{e}(h_{c.m.}) = 64\pi\epsilon \kappa R \left(\frac{k_B T}{e}\right)^2 \tanh{\left(\frac{e \Psi_{w}}{4 k_B T}\right)}   \nonumber \label{eq:model_e} \\ & \phantom{F_{e}(h_{c.m.}) =}\tanh{\left(\frac{e \Psi_{p}}{4 k_B T}\right)} e^{-\kappa h_{c.m.}} \\ 
    &F_g = -\frac{4}{3}\pi R^3 \left(\rho_p - \rho_f\right) g \label{eq:model_g}
\end{align} with $h_{c.m.}$ the height of the center of the sphere, $F_{e}(h_{c.m.})$ the force due to overlapping electric double layers of the particle and the wall, $F_g$ the gravitational force, $\epsilon$ the dielectric permittivity of water, $k_B$ the Boltzmann constant, $T=\SI{300}{\kelvin}$ the temperature, $e$ the elemental charge, $\Psi_p$ and $\Psi_w$ the Stern potentials of the particle and wall respectively, $\rho_p \approx \SI{2.0}{\gram\per\cubic\centi\meter}$ the particle density, $\rho_f$ the density of water, $g$ the gravitational acceleration and $\kappa^{-1}$ the Debye length. Based on the pH of our solution (pH$\approx$5.5), we find that the solution ionic strength is approximately $I = 10^{-5.5} = \SI{3e-6}{M}$. Therefore, the Debye length is expected to be $\kappa^{-1}(\si{nm}) = 0.304 / \sqrt{I(M)} = \SI{175}{\nm}$ \cite{israelachvili2011intermolecular}, in good agreement with the fit values of 100 to \SI{230}{\nm} that we obtained by fitting Equation 7 and Equation 11 to the experimental data for both sphere and dumbbell particles, respectively. We neglected van der Waals interactions; we used the Derjaguin approximations for $F_{e}$. For the electrostatic potential, we used the Debye-H{\"u}ckel approximation,
\begin{align}
    {\Psi}(r) = {\Psi_s}\frac{R}{r}\exp({-\kappa}(r-R)),\label{eq:stern}
\end{align} with $\Psi(r)$ the electrostatic potential at a distance $r$ from the center of the particle and $\Psi_s$ the Stern potential. By setting $\Psi(r=1/\kappa)$ equal to the here measured zeta potential of the spherical particles, see values in~\autoref{Sec:spheres}, we calculated an approximate value for the Stern potential. This we subsequently used as a starting value for the least-square fit of the model to our experimental height distributions. For the wall, we converted the zeta potential value of \SI{-55}{\milli\volt}~\cite{Gu2000} to an approximate Stern potential using \autoref{eq:stern}. For $\rho_p$ and $\Psi_{p}$ we used $\pm 2\sigma$ bounds; we fixed $\Psi_{w}$ to the calculated value as discussed above, and put no restrictions on $\kappa$.

To calculate the expected height distribution, we first obtained the electrostatic and gravitational potential energy, $\phi_{e}(h_{c.m.})$ and $\phi_{g}(h_{c.m.})$, respectively, from the force, 
\begin{align}
    &\phi_{e}(h_{c.m.}) = F_{e}(h_{c.m.})/\kappa \\ 
    &\phi_{g}(h_{c.m.}) = -F_{g} h_{c.m.},
\end{align} which we then used to derive the appropriate Boltzmann distribution, 
\begin{align}
    &p(h_{c.m.}) = A \exp\left(-\frac{\phi_{e}(h_{c.m.}) + \phi_{g}(h_{c.m.})}{k_B T}\right),\label{eq:p_sph}
\end{align} up to a normalization constant $A$~\cite{Wu2005}.

\subsection{Sphere Near-Wall Diffusion}

To test the validity of our measuring approach and the accuracy of our extracted gap heights above the wall, we sought to compare our measurements to theoretical predictions. To this end, and since well-established predictions exist for spheres alone, we determined the translational diffusion coefficient for our sphere measurements as function of gap height. To calculate the translational diffusion coefficient with gap height in~\autoref{Sec:spheres}, we proceeded as follows: instead of binning particle trajectories in time leading to bins with large height variations, we splitted all trajectories into shorter trajectories for which the gap height stayed within a certain height range, typically binning the total height range in bins of \SI{0.30}{\um} and \SI{0.12}{\um} for the 1.1 and \SI{2.1}{\um} spheres, respectively. For each height bin, the in-plane mean squared displacement (MSD, $\langle\Delta r^2\rangle$) was calculated. The in-plane translational diffusion coefficient $D$ and its error (standard deviation), was obtained from the first data point, typically an average of at least 300 measurements, of the MSD corresponding to a lag-time $\Delta t$ of 0.053 s using $\langle\Delta r^2\rangle = 4D\Delta t$.

\subsection{Modeling Forces and Torques on the Dumbbell}\label{Sec:dumb_model}

To elucidate dumbbell behaviors above the wall presented in~\autoref{sec:dumbbells_orient}, we extended the sphere model of~\autoref{Analysis:HeightDistribution} to our dumbbells. To this end, we approximated the gravitational and electrostatic forces acting on a dumbbell, by assuming that the spheres which comprise the dumbbell interact with the wall individually, as though the other is not present. That is, we use the expressions from~\Crefrange{eq:model_net}{eq:model_g} on each sphere, see \autoref{sec:theory} for the results. This approximation ignores the distortion of the electrostatic double layer caused by the presence of the other sphere, but allows us to derive predictions efficiently. We discuss the consequences of this approximation in \autoref{sec:theory}. The total force and torque acting on the dumbbell c.m. are thus given by:
\begin{align}
    F_{DB} &= F(h_1) + F(h_2) \label{eq:dumb_force} \\
    T_{DB} &= ((\mathbf{r}_1 - \mathbf{r}_{c.m.}) \cross F(h_1) \mathbf{\hat{e}}_z \nonumber \\ &\phantom{=}+ (\mathbf{r}_2 - \mathbf{r}_{c.m.}) \cross F(h_2) \mathbf{\hat{e}}_z) \cdot \mathbf{\hat{e}}_x \label{eq:dumb_torque} 
\end{align} with $h_i, \mathbf{r}_i$ the height and position of sphere $i$, $\theta_p$ the angle between the long axis of the dumbbell and the wall and $\mathbf{\hat{e}}_j$ the unit vector along the $j \in [x, y, z]$ axis (see \autoref{fig:fig1}b for a schematic).

From the force expressions acting on the individual spheres of the dumbbell, we calculated the corresponding potential energy:
\begin{align}
    &\phi_{DB}(h_{c.m.}, \theta_p) = -2 F_{g} h_{c.m.} + \frac{2 F_{e}(h_{c.m.})}{\kappa} \cosh{\left(\kappa R \sin{\theta_p}\right)}. \label{eq:dumb_pot}
\end{align} 
\autoref{eq:dumb_pot} assumes both spheres to have the same radius, see \autoref{eq:uneqphi} for a general expression for dumbbells made of spheres of unequal radii.
This potential can be derived with respect to the $\mathrm{h_{c.m.}}$ to obtain the force and to $\theta_p$ to obtain the torque. We subsequently used the potential to derive the appropriate height distribution for the dumbbell c.m. $p_{DB}(h_{c.m.}, \theta_p)$ up to a normalization constant,
\begin{align}
    &p_{DB}(h_{c.m.}, \theta_p) \propto K\exp\left[-\frac{\phi_{DB}(h_{c.m.}, \theta_p)}{k_B T}\right] \label{eq:db_prob}\\
    &p_{DB}(h_{c.m.}) \propto \int^{\frac{\pi}{2}}_{-\frac{\pi}{2}} d\theta_p \cos{(\theta_p)} K\exp\left[-\frac{\phi_{DB}(h_{c.m.}, \theta_p)}{k_B T}\right],\label{eq:db_height} \\
    &p_{DB}(\theta_p) \propto \int^{h_{max}}_{R} dh_{c.m.} K\exp\left[-\frac{\phi_{DB}(h_{c.m.}, \theta_p)}{k_B T}\right],\label{eq:db_angle}
\end{align}
where we evaluated \autoref{eq:db_height} by numeric integration over all possible plane angles $\theta_p$, and \autoref{eq:db_angle} by numeric integration over all possible heights $h_{c.m.}$; $h_{max}$ was set to \SI{5}{\um}. $K$ represents the particle-wall hard-core interaction potential contribution to the Boltzmann weight: $K=1$ if both spheres of the dumbbell are above the wall; otherwise $K=0$. We have calculated the probability as function of the lowest dumbbell gap height (i.e., the separation between the wall and the bottom of the lower sphere of the dumbbell) by substituting $h_{c.m.} = h_{g, l}+R+R\sin{\theta_p}$ in \autoref{eq:db_height}. Equivalently, for the upper gap height, we substituted $h_{c.m.} = h_{g, u}+R-R\sin{\theta_p}$ in \autoref{eq:db_height} to derive its distribution.

\section{Results and Discussion}

\subsection{Characterization, Height Distribution, and Diffusion with Wall Gap Height, of Spherical Particles above the Wall}\label{Sec:spheres}

\begin{figure}
    \centering
    \includegraphics{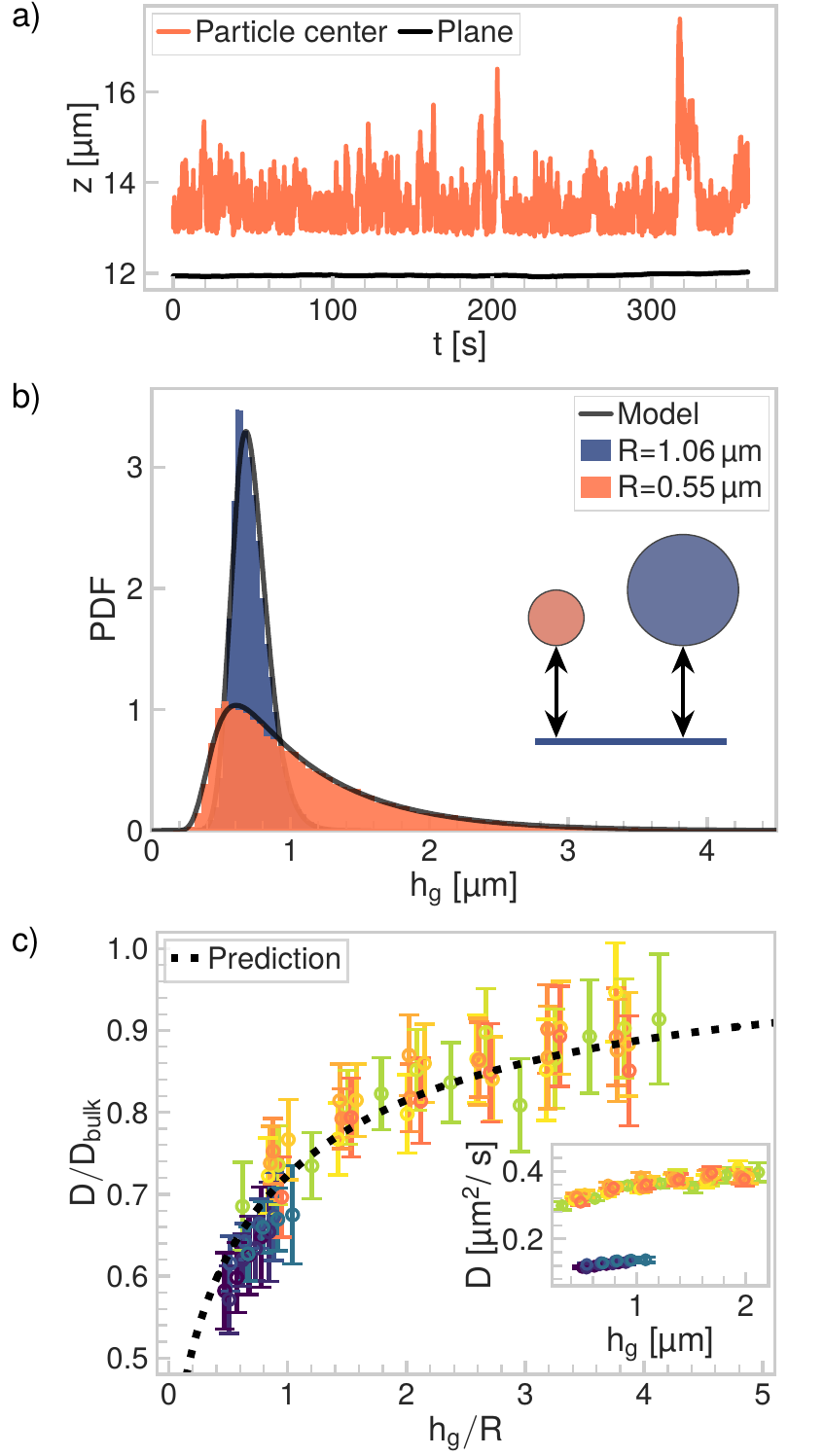}
    \caption{\textbf{Sphere-wall gap height and translational diffusion above a planar wall.} \textbf{a)} The z positions of a spherical particle diffusing above a wall, as well as that of the plane (directly below the particle) obtained from the positions of three spheres fixed on the wall, are plotted in time. Using the plane z position, the gap height $h_g$ between the diffusive particle and wall is determined. \textbf{b)} Experimental sphere-wall gap height distributions together with a fit with the model from Ref.~\cite{Wu2005} which combines gravitational and electrostatic effects for \SI{1.1}{\um} (orange, fit parameters $\rho_p=\SI{2.1}{\gram\per\cubic\centi\meter}$, $1/\kappa=\SI{107}{\nm}$, $\zeta_p=\SI{-41}{\milli\volt}$) and \SI{2.1}{\um} (blue, fit parameters $\rho_p=\SI{2.2}{\gram\per\cubic\centi\meter}$, $1/\kappa=\SI{207}{\nm}$, $\zeta_p=\SI{-52}{\milli\volt}$) spheres. \textbf{c)} Normalized translational near-wall in plane diffusion coefficient of \SI{1.1}{\um} (light) and \SI{2.1}{\um} (dark) spheres as function of normalized gap height. Error bars denote standard deviations. Experimental data are plotted against the theoretical prediction that follows from Ref.~\cite{Ketzetzi2020Arxiv}. Inset shows the non-normalized diffusion coefficient values for both sphere sizes with gap height.} 
    \label{fig:fig2}
\end{figure}

First, we measured the sphere dynamics above a planar wall both to assess the sensitivity of our LED-based in-line holographic microscopy setup, as well as to verify our new method of using fixed particles to accurately locate the position of the wall. Indeed, despite the simplicity of our setup, we find an excellent agreement between the measured holograms and the Mie scattering-based model, see~\autoref{fig:fig1}d for a direct comparison that additionally shows the residual between data and model. Moreover, in steps 2 and 3 of~\autoref{fig:fig1}e we show the refractive indices and particle radii that we obtained during characterization, respectively. Both parameters agree with expectations: the refractive index, $n_{\mathrm{silica}} = (1.42\pm0.02)$ agrees with the value provided by the supplier ($1.42$) and at the same time the radius of the particles ($\SI{0.51\pm0.03}{\um}$) follows our TEM results ($\SI{0.48\pm0.03}{\um}$). 

For high precision measurements, careful consideration should be given to the determination of both the position and local orientation of the wall, from which the gap height can be derived, as walls in experiments may be tilted. Here, we achieved such precision (see \autoref{fig:fig2}a), by using at least three fixed particles that define a plane and by subsequently obtaining the position of the diffusing particle relative to said plane. Note that the position and orientation of the plane is fitted accurately to the positions of the bottom of the fixed particles, since our method also measures the radii of the fixed particles at the same time.

In \autoref{fig:fig2}b, we report the distribution of gap heights between the diffusing spheres of two different sizes and the wall. We find that the height distributions can faithfully be described using established methods that combine a barometric height distribution with electrostatic interactions (see also~\autoref{Analysis:HeightDistribution} and Ref.~\cite{Wu2005}). In comparison, the height distributions of the \SI{1.1}{\um} and \SI{2.1}{\um} spheres feature qualitatively different behaviors. As expected, the smaller spheres probe a wider range of gap heights, while the axial motion of the larger spheres is more confined. However, we note that the median gap height of the larger spheres is slightly greater than that of the smaller ones, which is in line with the higher surface charge that we measured for these particles using laser doppler micro-electrophoresis. The corresponding zeta potentials are \SI{-54 \pm 7}{mV} and \SI{-35 \pm 6}{mV} for the 1.1 and \SI{2.1}{\um} batches, respectively. The excellent agreement that we obtained between the prediction and our experiment for different particle parameters further verifies the sensitivity of our setup. We conclude that our method of localizing the plane, and thereby the wall, using fixed control particles allows for high precision measurements of colloidal systems near walls.

Finally, to further evaluate our method, we determined the height-dependent translational diffusivity of the spheres, presented in~\autoref{fig:fig2}c. Additionally, in the same figure, we compared our data to the theoretical prediction for translational diffusion with wall gap height of Ref.~\cite{Ketzetzi2020Arxiv}, which covers the entire separation range from the far-field regime captured by Fax\'en~\cite{Faxen1921} and the near-wall regime captured by lubrication theory~\cite{Goldman1967}. We find that both particle sizes follow the prediction within error, with small random variations between individual measurements, which demonstrates that we can accurately determine the diffusion constant across the whole range of here accessible sphere-wall gap heights.

\subsection{Dumbbell Height Distribution Above the Wall}\label{sec:dumbbells_height}

\begin{figure*}
    \centering
    \includegraphics{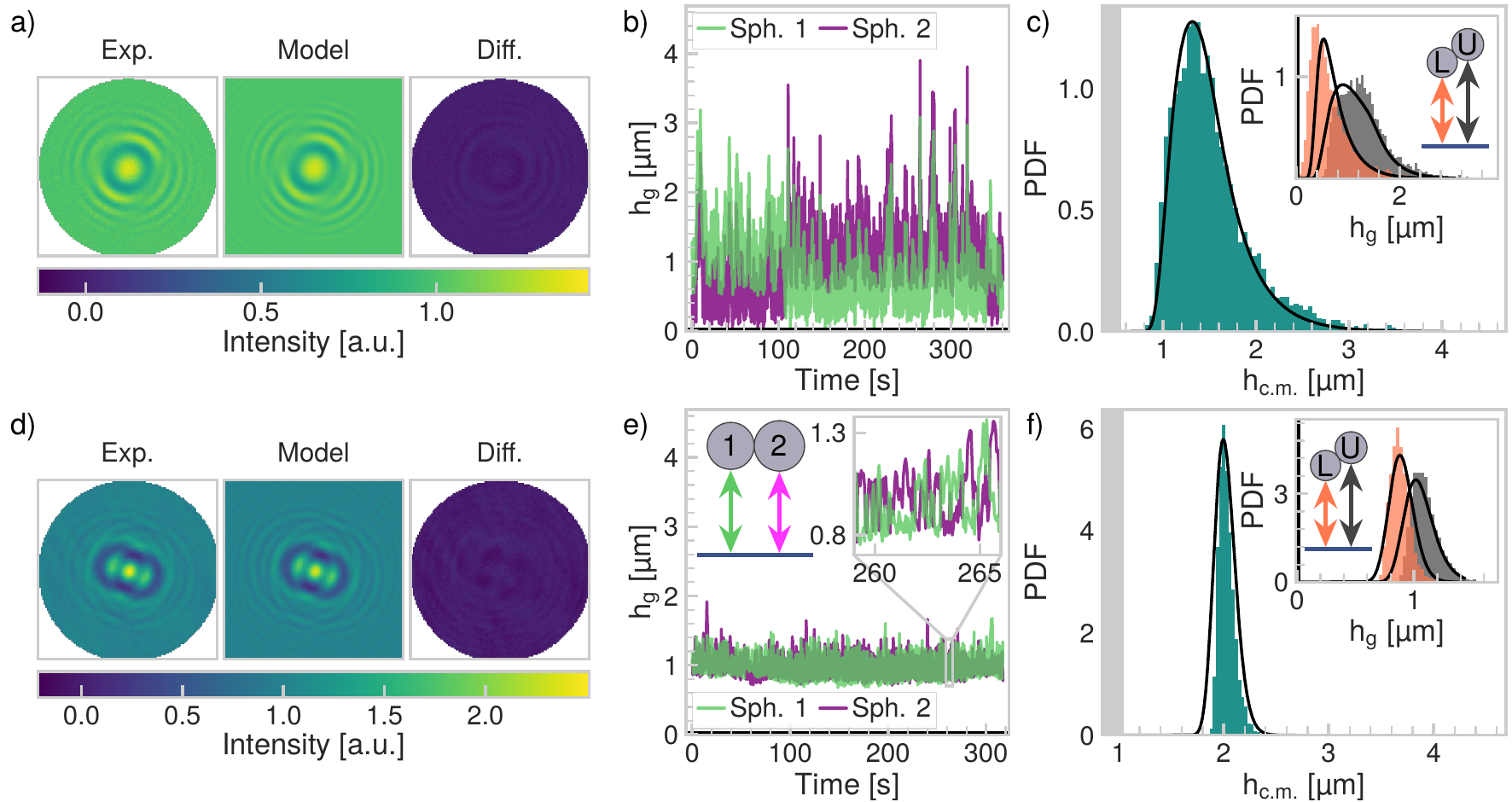}
    \caption{\textbf{Height distributions of colloidal dumbbells above a planar wall.} \textbf{a)} Comparison of an experimental image, the fitted model and the residual for a \SI{2.2}{\um} dumbbell, the low values of which indicate the good agreement between experimental data and model. \textbf{b)} Gap heights for the two \SI{1.1}{\um} spheres that form the dumbbell as function of time. \textbf{c)} Center of mass (c.m.) dumbbell height distributions (same particle as in b), with the corresponding gap heights of the lower (L) and upper (U) spheres as inset. Solid lines indicate the theoretical prediction of \autoref{eq:db_height} (fit parameters $\rho_p=\SI{2.0}{\gram\per\cubic\centi\meter}$, $1/\kappa=\SI{103}{\nm}$, $\zeta_p=\SI{-30}{\milli\volt}$). \textbf{d)} Comparison of an experimental image, the fitted model and the residual for a \SI{4.2}{\um} dumbbell, the low values of which indicate the excellent agreement between data and model. \textbf{e)} Gap heights for the two touching \SI{2.1}{\um} spheres that form the dumbbell as function of time. The inset zooms in on a short sequence of the measurement to indicate the frequent flipping of the dumbbell. \textbf{f)} Center of mass (c.m.) dumbbell height distributions (same particle as in e), with the corresponding dumbbell gap heights of the lower (L) and upper (U) spheres as inset. Solid lines indicate the theoretical prediction of \autoref{eq:db_height} (fit parameters $\rho_p=\SI{2.1}{\gram\per\cubic\centi\meter}$, $1/\kappa=\SI{228}{\nm}$, $\zeta_p=\SI{-61}{\milli\volt}$).}
    \label{fig:fig3}
\end{figure*}

Having established the validity of our setup and method, we proceeded to study the near-wall behavior of our colloidal dumbbells. These dumbbells were formed by random aggregation of two individual spheres caused by Van der Waals attraction; we expect that the spheres do not roll with respect to each other. Analogously to the spheres, we measured the three-dimensional position of dumbbells of two sizes (long axis 2.2 and 4.2 \si{\um} respectively), formed either by two 1.1 \si{\um} or two 2.1 \si{\um} spheres. We first checked the quality of our hologram analysis in~\autoref{fig:fig3}a and d, where the good agreement between the model and our experimental images is shown. In this model, the free parameters are the c.m. position, the dumbbell orientation, the radii, and the refractive indices of the two touching spheres comprising the dumbbell. We note that the obtained values agreed with the single spheres results (\autoref{fig:fig1}e step 2 and 3).

\autoref{fig:fig3}b shows the positions of the \SI{1.1}{\um} spheres comprising the dumbbell (dumbbell long axis \SI{2.2}{\um}) as function of time, revealing that one of the spheres is positioned higher than the other in relation to the wall. Moreover, it clearly shows that twice during the duration of our 8 min measurement, the spheres drastically changed positions, \textit{i.e.}, a flipping between upper and lower spheres occurred. Based on the estimated rotational diffusion time $\tau_r = 8\pi\eta R_{eff}^3 / (k_B T) \approx \SI{2}{\second}$ (with viscosity $\eta = \SI{8.9e-4}{\pascal\second}$ and the radius of a sphere of volume equal to the dumbbell $R_{eff} = (2 R^3)^{1/3} \approx \SI{0.69}{\um}$), this flipping should have been observed more frequently if it were a purely diffusive process, faraway from the wall. For the larger dumbbells in~\autoref{fig:fig3}e, which move further from the wall, we observe despite their larger size ($\tau_r \approx \SI{13}{\second}$), frequent flipping between the upper and lower spheres.

By fitting the c.m. height distribution of the dumbbell in~\autoref{fig:fig3}c and f using \autoref{eq:db_height} (solid black line), we conclude that our simple model for a dumbbell particle near a wall describes the experimental height distribution very well. Furthermore, the fit parameters we have obtained from this fit agree with the single sphere fit parameters from the height distribution in \autoref{fig:fig2}b. Additionally, we calculate the height distribution of the dumbbell gap heights of the lower (L) and upper (U) spheres, as shown in the inset of \autoref{fig:fig3}c and f. Compared to the theoretical prediction from \autoref{eq:db_height}, we observe a slight shift towards smaller heights for the lower and, conversely, greater heights for the upper sphere in the experiments. This may indicate that to fully describe the experimental data, higher order effects need to be taken into account, such as the distortion of the electric double layer of one sphere by the presence of the other sphere and the wall. These effects become more pronounced when the dumbbells are closer to the wall, as can be seen when comparing panels c and f from \autoref{fig:fig3}.

\subsection{Dumbbell Orientation with Respect to the Wall}\label{sec:dumbbells_orient}

\begin{figure*}
    \centering
    \includegraphics{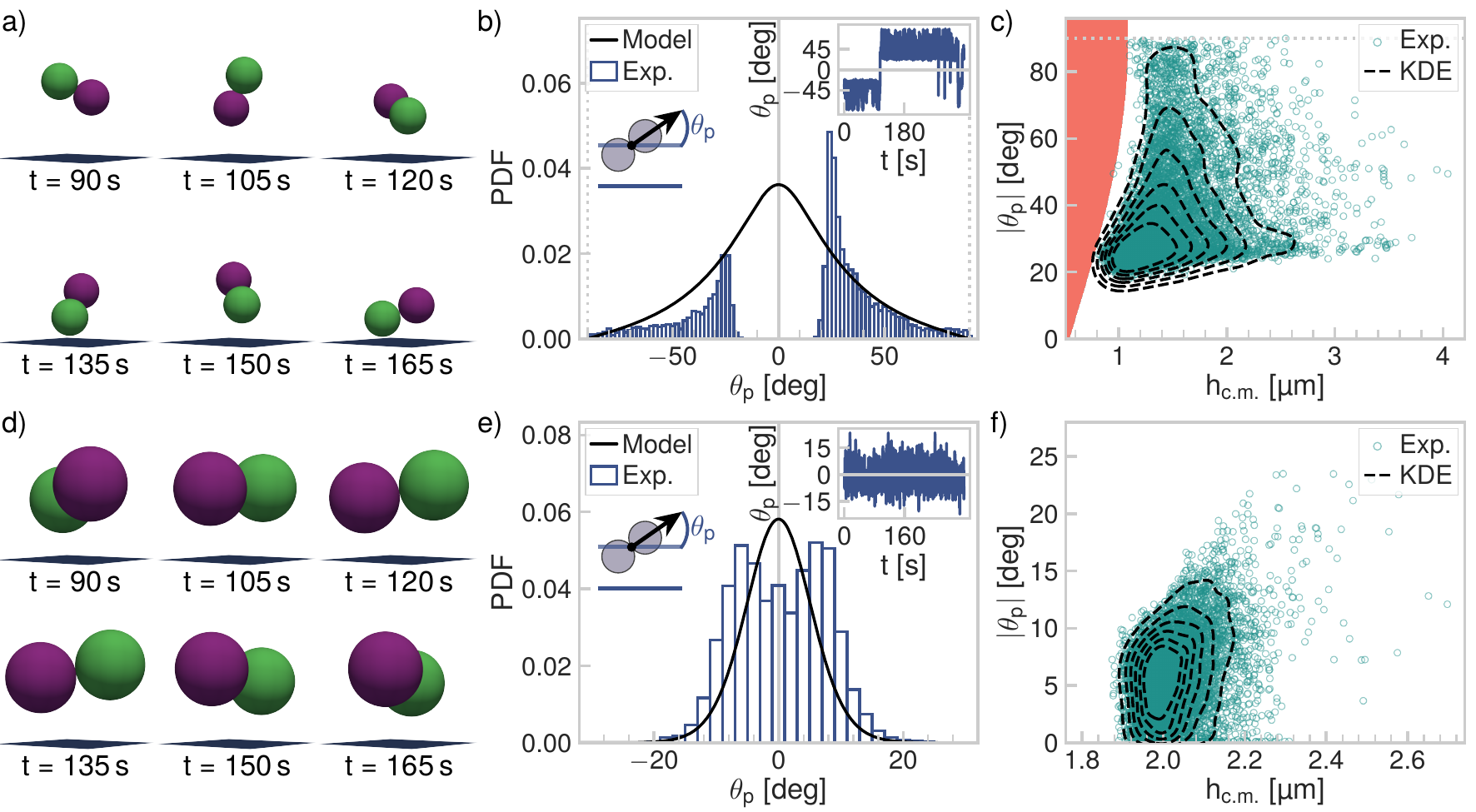}
    \caption{\textbf{Dumbbell orientation with respect to the planar wall as function of height.} \textbf{a)} Schematics based on the experimentally tracked positions of a \SI{2.2}{\um} dumbbell at random times, showing out of plane rotations in addition to height variations. \textbf{b)} Distribution of plane angles for a \SI{2.2}{\um} dumbbell. The difference in peak heights is due to the respective length of the parts of the measurement where the dumbbell assumed a negative or positive orientation (see inset). We distinguish negative from positive orientations as outlined in \autoref{Analysis:Dumbbells}. The solid line indicates the expected distribution based on \autoref{eq:db_angle} (same parameters as in \autoref{fig:fig3}c). The inset shows the plane angle in time. \textbf{c)} Plane angle with c.m. height for the \SI{2.2}{\um} dumbbell. The red area indicates geometrically forbidden configurations. \textbf{d)}  Schematics based on the experimentally tracked positions of a \SI{4.2}{\um} dumbbell at the same times as in (a), showing significantly fewer out of plane rotations compared to the smaller dumbbell of (a). \textbf{e)} Distribution of plane angles for a \SI{4.2}{\um} dumbbell. The solid line indicates the expected distribution based on \autoref{eq:db_angle} (same parameters as in \autoref{fig:fig3}f). The inset shows the plane angle in time. \textbf{f)} Plane angle with c.m. height for the \SI{4.2}{\um} dumbbell. In panel c and f, the dashed lines are a contour plot of the kernel density estimation, corresponding to \SI{12.5}{\percent}, \SI{25}{\percent}, \SI{37.5}{\percent}, \SI{50}{\percent}, \SI{62.5}{\percent} and \SI{75}{\percent} of the data.} 
    \label{fig:fig4}
\end{figure*}

The stable and significant differences in sphere positions of \autoref{fig:fig3}b, already indicated that these dumbbells are oriented at an angle relative to the wall. On the other hand, for larger dumbbells of the same material, the spheres being approximately at the same height at all times in~\autoref{fig:fig3}e suggested a roughly parallel orientation with the wall. We verify our observations in~\autoref{fig:fig4}a and~\ref{fig:fig4}d, where we visualize orientations that the dumbbells assumed during the measurements at \SI{15}{\second} intervals. Indeed, from the snapshots we clearly see that, while flipping between lower and upper sphere did occur, the height above as well as orientation with respect to the wall remained relatively constant for the larger dumbbell~(\autoref{fig:fig4}d). Conversely, the smaller dumbbell featured a richer behavior that includes notable changes in height, orientation, as well as flipping between which of the two spheres is the lowest~(\autoref{fig:fig4}a).

In what follows, we further quantify our observations, by calculating the angle, $\theta_p$, between the long dumbbell axis and wall (see schematic of \autoref{fig:fig4}b). Strikingly, we observe in \autoref{fig:fig4}b a double-peaked structure not predicted by our model: we find no parallel orientations with respect to the wall for the the \SI{2.2}{\um} dumbbell. Instead, the dumbbell is more likely to be oriented at an angle between 25 and \SI{56}{deg} (median \SI{32}{deg}) with the wall. In separate bright-field microscopy measurements, we verified that dumbbells of this size and material indeed show frequent out-of-plane rotations. The preferred range of orientations is robust, and persists even when the dumbbell flips, i.e. when the lower sphere becomes the upper sphere. The difference in peak heights in~\autoref{fig:fig4}b is due to the respective length of the parts of the measurement where the dumbbell assumed a negative or positive orientation. Such preferred orientations are surprising, since an angle distribution centered around zero degrees is naively expected in view of the effects of buoyancy and electrostatics, see the expected distribution depicted by the solid line in~\autoref{fig:fig4}b and~\ref{fig:fig4}e. 

Examining the larger and hence heavier \SI{4.2}{\um} dumbbells in~\autoref{fig:fig4}e, we notice that these indeed have assumed mostly flat orientations with the wall, with the most probable angles ranging between 2.2 and \SI{9.6}{deg} (median \SI{6}{deg}). However, the double-peak structure in the angle probability density function that we observed for the smaller dumbbells persists to some degree even for these larger particles, indicating that the increased gravitational force leads to a suppression of the interaction which causes the dumbbells to adopt a nonparallel orientation. We hypothesize that the observed angle distributions for both dumbbell sizes stem from a higher-order electrostatic effect not accounted for in our theory. However, we cannot exclude a more subtle interplay of other effects, such as buoyancy and hydrodynamics.

Naturally, the question arises whether changes in height relate to changes in dumbbell orientation. To test for this, we plot the measured angles as function of center-of-mass height. We find that for the smaller dumbbells, there is a clear preference for lower angles at low heights in~\autoref{fig:fig4}c, the preference for which disappears with height. That is, further from the wall, the dumbbells may adopt a wider range of orientations. For the larger dumbbell, we also find a narrower distribution of angles at lower heights in~\autoref{fig:fig4}f. However, we note that both angle and height distributions are considerably narrower compared to those that correspond to the smaller dumbbell. At the same time, the particle-wall separation distance is typically greater than that of the smaller dumbbell: while the smaller dumbbell moves closely to the wall (see also the red area in \autoref{fig:fig4}c which indicates geometrically forbidden configurations caused by particle-wall overlap), the larger dumbbell does not come into close contact with the wall.

\subsection{Theoretical Considerations for Preferred Dumbbell Orientations}\label{sec:theory}

\begin{figure*}
   \centering
   \includegraphics{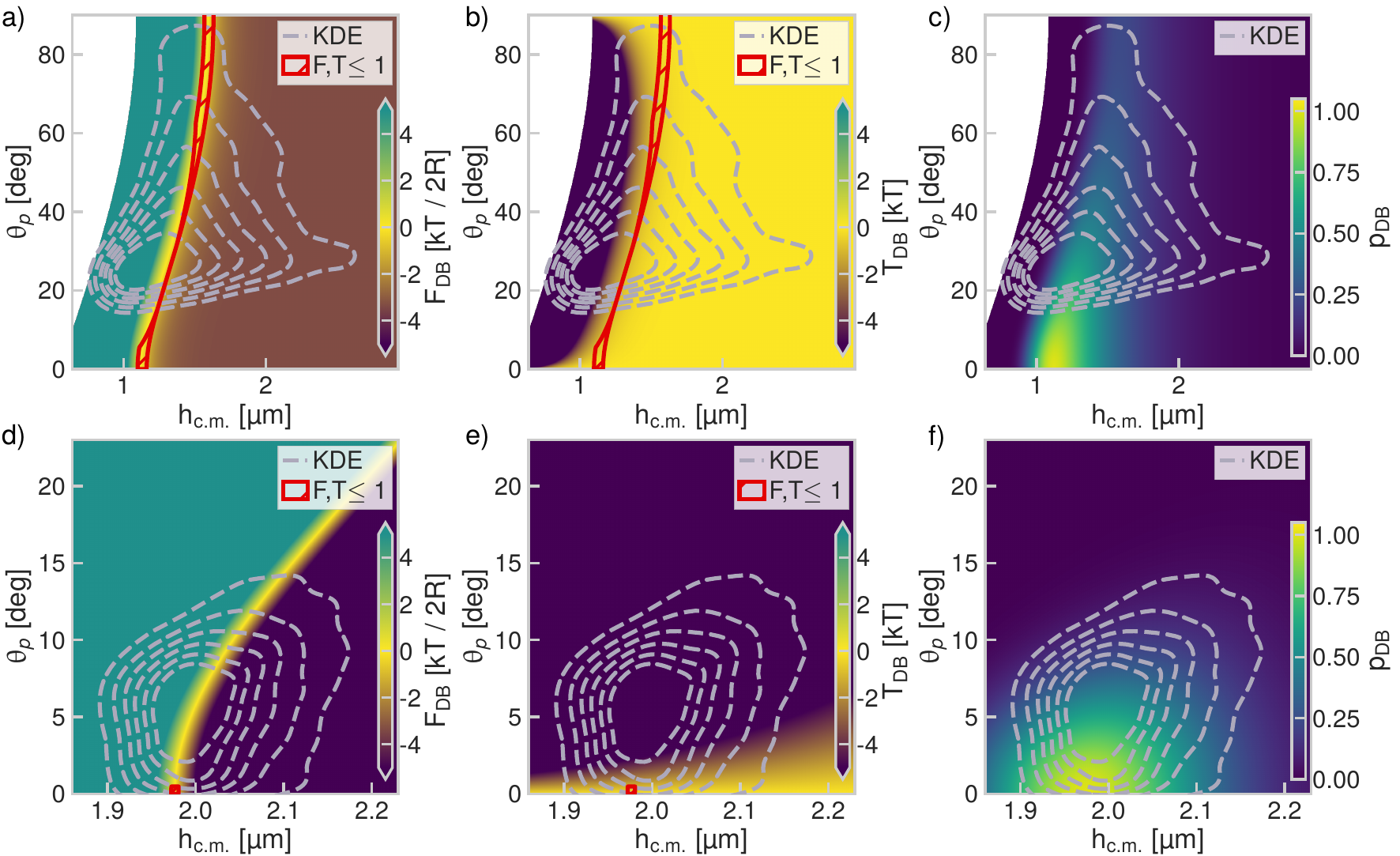}
   \caption{\textbf{Force and torque acting on a dumbbell by balancing electrostatics and gravity.}
   \textbf{a)} The force as function of $\theta_p$ and $h_{\mathrm{c.m.}}$ for the \SI{2.2}{\um} dumbbell. For all orientations, there is a height range for which the net force is zero. \textbf{b)} The torque as function of $\theta_p$ and $h_{\mathrm{c.m.}}$ for the \SI{2.2}{\um} dumbbell. \textbf{c)} The probability of observing a combination of $\theta_p$ and $h_{\mathrm{c.m.}}$ for the \SI{2.2}{\um} dumbbell, as predicted by \autoref{eq:db_prob} and measured in the experiments (dashed line). \textbf{d)} The force as function of $\theta_p$ and $h_{\mathrm{c.m.}}$ for the \SI{4.2}{\um} dumbbell. The area where the net force is zero is smaller compared to the smaller dumbbell in (a). \textbf{e)} The torque as function of $\theta_p$ and $h_{\mathrm{c.m.}}$ for the \SI{4.2}{\um} dumbbell. For the same range of angles as in (b), the torque on the larger dumbbell is considerably higher than the thermal energy for the majority of angles, causing the dumbbell to adopt a flat orientation with respect to the wall. \textbf{f)} The probability of observing a combination of $\theta_p$ and $h_{\mathrm{c.m.}}$ for the \SI{4.2}{\um} dumbbell. In panels a, b, d and e, the red lines indicate regions where both the force and torque are simultaneously small compared to the thermal energy, indicating a possibility of observing the dumbbell at those heights and orientations. Values outside the indicated range of the color-bars are clipped to visualize the low force and torque region relevant to the experiments, while white regions represent sterically forbidden combinations of height and angle. Dashed lines are a contour plot of the kernel density estimation of the experimental data (see \autoref{fig:fig4}). \label{fig:fig5}}
\end{figure*}

To gain insight into the preferred orientations and minimal angle measured in \autoref{sec:dumbbells_orient}, we extended the gravity and electrostatics model for a sphere above the wall (\Crefrange{eq:model_net}{eq:model_g}) to the dumbbell. Briefly, \Crefrange{eq:dumb_force}{eq:dumb_torque} model the dumbbell as two connected (but otherwise non-interacting) spheres, by balancing electrostatic and gravitational forces. This approximation ignores the distortion of the electrostatic double layer caused by the presence of the other sphere, but allowed us to probe the origin of the dumbbell orientation described in~\autoref{sec:dumbbells_orient}, by examining whether the combined effects of electrostatics and gravity would result in zero force and torque solutions as function of plane angle and height above the wall. 

By applying the reduced model of \autoref{Sec:dumb_model} to the experimental data, we reach a number of interesting conclusions in~\autoref{fig:fig5}, where we plot the results from the model. \autoref{fig:fig5}a shows that the net force on the \SI{2.2}{\um} dumbbell vanishes for a range of heights and orientations. That is, for each given orientation there exists a narrow distribution of heights where the force balance is zero. As expected for a particle with a larger mass, for the \SI{4.2}{\um} dumbbell in~\autoref{fig:fig5}d the range of heights where the net force vanishes is considerably narrower compared to the \SI{2.2}{\um} dumbbell of~\autoref{fig:fig5}a. To answer whether such configurations are expected to be stable, one must additionally consider the possibility of a reorienting torque stemming from the combined effect of gravity and electrostatics acting on the dumbbell.
We expect that the interplay between the magnitude of this reorienting torque and a random torque, stemming from thermal fluctuations, causes changes in the dumbbell orientations with respect to the wall. In the case of a reorienting torque that is large in comparison to the random torque ($\approx$~1~kT), we expect a mostly parallel orientation with respect to the wall. In contrast, for a reorienting torque that is small compared to the random torque, we expect largely fluctuating orientations. In what follows, we examine the presence and magnitude of the reorienting torque.

Interestingly, for the smaller \SI{2.2}{\um} dumbbells, a regime arises where both net forces and reorienting torques are simultaneously below the thermal force and energy, respectively, for certain combinations of dumbbell-wall separations and non-zero plane angles (as indicated by the red lines in \autoref{fig:fig5}a and b). The presence of such a regime that spans throughout state space suggests that the large variations of the angle as found in \autoref{fig:fig4} (evidenced also in the dashed lines of \autoref{fig:fig5}a and b) are expected.  
This is further corroborated by the angle probability plot that follows from our model in \autoref{fig:fig5}c for heights relevant to our experiment. For the largest dumbbells, our minimal modeling (\autoref{fig:fig5}d-f) agrees well with the almost parallel orientations observed in the experiments (\autoref{fig:fig4}f), which mostly fall within the high reorienting torque regime (see dashed line in \autoref{fig:fig5}d).

Our minimal dumbbell model also sheds light on the relation between height and orientation observed in~\autoref{fig:fig4}c and f, indicated also by the dashed lines in~\autoref{fig:fig5}. Although the agreement is not fully quantitative, the model shown in \autoref{fig:fig5}c and f predicts an increase in the most probable angle with greater heights, similar to our experiments. Moreover, the height and orientation combinations that the dumbbells experimentally adopt most often coincide with the zero net force regime (and equivalently non-zero probabilities in \autoref{fig:fig5}c and f) for both dumbbell sizes, as evidenced by the overlap between the experimental data and the areas of higher probability.

Finally, we notice that the range of experimentally observed angles for the \SI{2.2}{\um} dumbbells does not fully coincide with the range of angles that fall within the low force and torque regime from the model. For torques below the thermal energy, the model also allows for angles below \SI{17}{deg}, which we did not observe here for these dumbbells. We note that the discrepancy between our model and experiment does not stem from a difference in size between the two spheres in the dumbbell. As can be seen in \autoref{fig:appendix1} and \autoref{fig:appendix2} where we additionally account for (an experimentally relevant) \SI{5}{\percent} dispersity in the sphere sizes, the most probable heights are only slightly shifted towards greater values. However, the overall dumbbell behavior that the model yields remains the same with or without polydispersity in the sphere size. We hypothesize that this discrepancy may be resolved by considering higher-order electrostatic effects. However, higher-order effects, together with the possibility of dynamic charge redistribution in the double layers which may be relevant here, cannot be described by a simple analytical model.

\section{Summary and Conclusion}

We have measured the height of colloidal particles relative to planar walls with high precision by means of holographic microscopy. The position of the wall was tracked in time by following the position of spheres fixed on its surface, thereby allowing for an accurate measurement of the location and orientation of the plane and wall. For spheres, the obtained height distributions and diffusivities as function of height are in line with well-known theoretical predictions. 
More importantly, we studied the height distributions and orientations of colloidal dumbbells relative to walls. We found that smaller dumbbells assume non-parallel orientations with the wall and further examined the connection between orientation and particle-wall separation. Conversely, we found that larger dumbbells of the same material were always oriented almost parallel to the wall. 

We showed that, despite its simplicity, a minimal model accounting for gravity and electrostatics not only faithfully describes the dumbbell height distribution, but also predicts stable configurations for a large range of orientations and dumbbell-wall separations. However, our model predicts a larger range of stable orientations than was found in our experiment, indicating that refinements that account for higher-order electrostatic effects may need to be considered. We thus hope that our findings will encourage further investigations of near-wall particle dynamics. Our results highlight the rich dynamics that nonspherical particles exhibit in the proximity of walls and can aid in developing quantitative frameworks for arbitrarily-shaped particle dynamics in confinement.
\\

We gratefully acknowledge Samia Ouhajji for providing the \SI{1.1}{\um} silica spheres. We thank Sarah Smolders for exploratory experiments, and Nikos Oikonomeas for exploratory experiments and useful discussions on analyzing digital holograms. J.d.G. thanks NWO for funding through Start-Up Grant 740.018.013 and through association with the EU-FET project NANOPHLOW (766972) within Horizon 2020. D.J.K. gratefully acknowledges funding from the European Research Council (ERC) under the European Union's Horizon 2020 research and innovation program (grant agreement no. 758383). 

\section{Colloidal dumbbells of differently sized spheres: plane height and orientation probability density}\label{sec:appendix}

\begin{figure*}
    \centering
    \includegraphics[width=\linewidth]{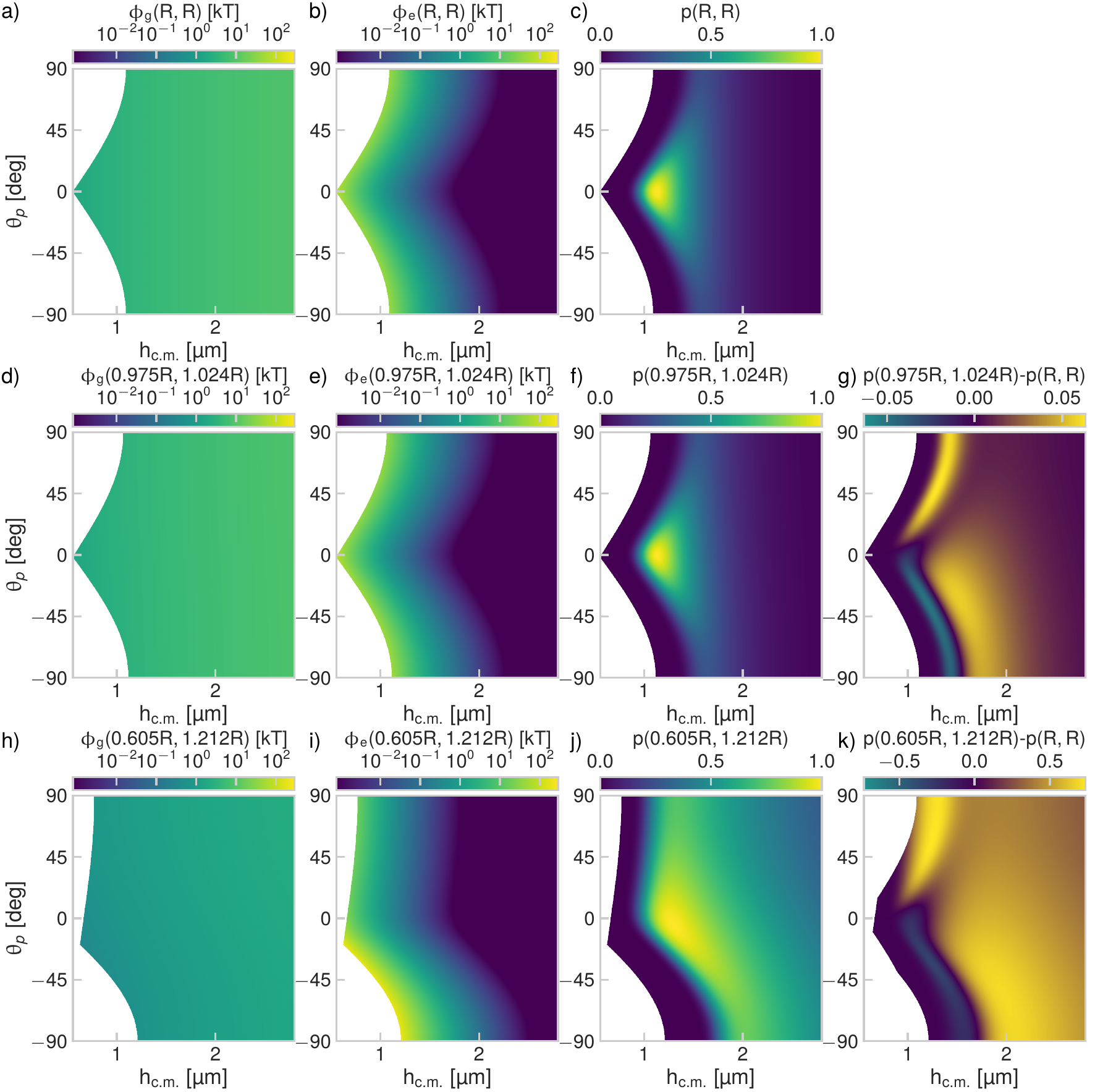}
    \caption{\textbf{Effect of sphere size dispersity on dumbbell plane height and orientation probability density for dumbbells of the same volume as the \SI{2.2}{\um} dumbbells ($R=\SI{0.54}{\um}$).} All gravitational and electrostatic potentials were calculated according to \autoref{eq:uneqphig} and \autoref{eq:uneqphie}, respectively. All probabilities were calculated according to \autoref{eq:uneqp}. \textbf{a)} Gravitational potential for $R_1=R_2=R$. \textbf{b)} Electrostatic potential for $R_1=R_2=R$. \textbf{c)} PDF for $R_1=R_2=R$. \textbf{d)} Gravitational potential for $R_1=0.975R, R_2=1.024R$. \textbf{e)} Electrostatic potential for $R_1=0.975R, R_2=1.024R$. \textbf{f)} PDF for $R_1=0.975R, R_2=1.024R$. \textbf{g)} Probability difference $p(0.975R, 1.024R) - p(R, R)$. \textbf{h)} Gravitational potential for $R_1=0.605R, R_2=1.212R$. \textbf{i)} Electrostatic potential for $R_1=0.605R, R_2=1.212R$. \textbf{j)} PDF for $R_1=0.605R, R_2=1.212R$. \textbf{k)} Probability difference $p(0.605R, 1.212R) - p(R, R)$.\label{fig:appendix1}}
\end{figure*}

\begin{figure*}
    \centering
    \includegraphics[width=\linewidth]{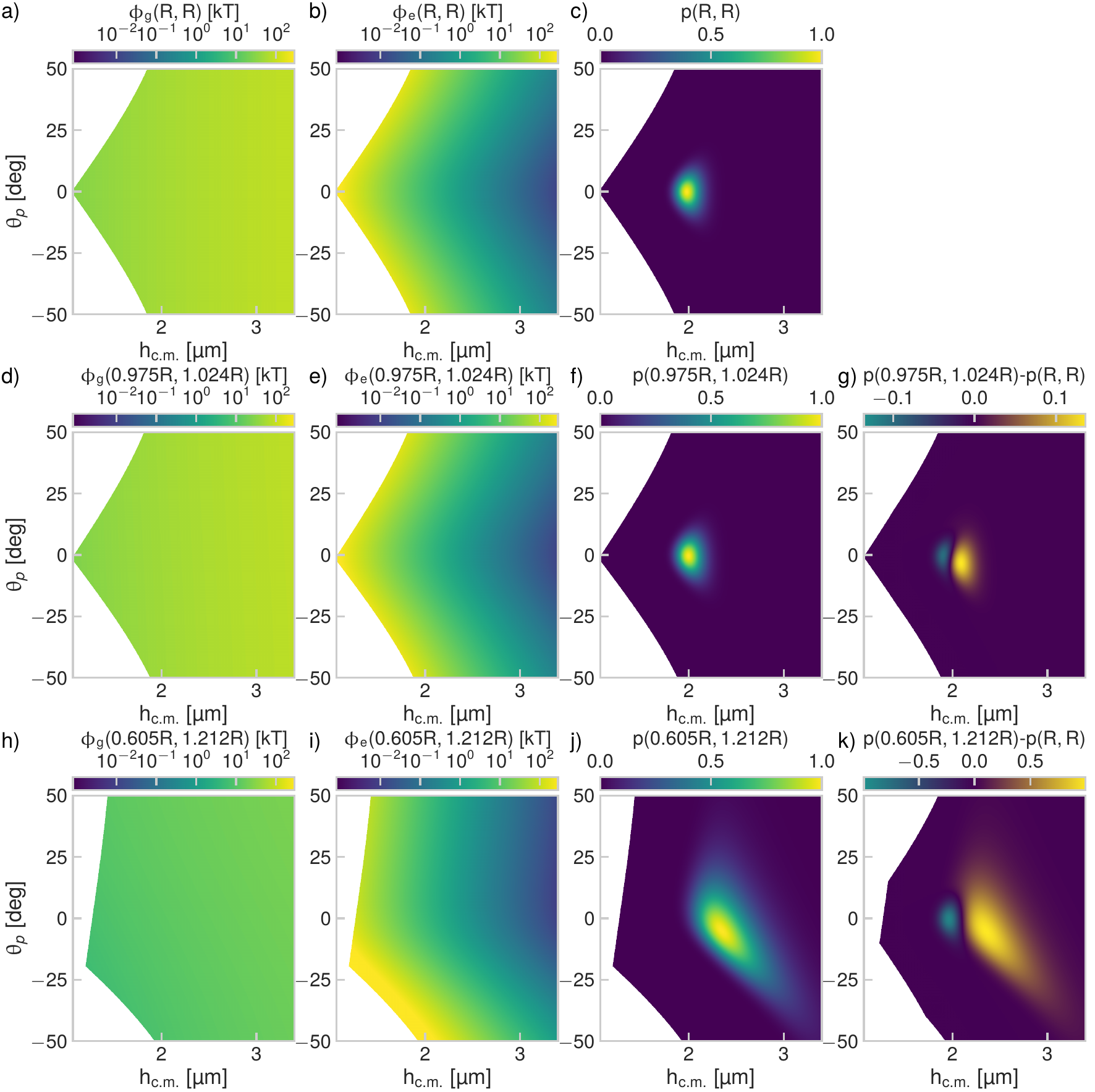}
    \caption{\textbf{Effect of sphere size dispersity on dumbbell plane height and orientation probability density for dumbbells of the same volume as the \SI{4.2}{\um} dumbbells ($R=\SI{1.04}{\um}$).} All gravitational and electrostatic potentials were calculated according to \autoref{eq:uneqphig} and \autoref{eq:uneqphie}, respectively. All probabilities were calculated according to \autoref{eq:uneqp}. \textbf{a)} Gravitational potential for $R_1=R_2=R$. \textbf{b)} Electrostatic potential for $R_1=R_2=R$. \textbf{c)} PDF for $R_1=R_2=R$. \textbf{d)} Gravitational potential for $R_1=0.975R, R_2=1.024R$. \textbf{e)} Electrostatic potential for $R_1=0.975R, R_2=1.024R$. \textbf{f)} PDF for $R_1=0.975R, R_2=1.024R$. \textbf{g)} Probability difference $p(0.975R, 1.024R) - p(R, R)$. \textbf{h)} Gravitational potential for $R_1=0.605R, R_2=1.212R$. \textbf{i)} Electrostatic potential for $R_1=0.605R, R_2=1.212R$. \textbf{j)} PDF for $R_1=0.605R, R_2=1.212R$. \textbf{k)} Probability difference $p(0.605R, 1.212R) - p(R, R)$.\label{fig:appendix2}}
\end{figure*}

Here we derive the electrostatic and gravitational forces on a dumbbell of two unequally sized spheres of radii $R = R_1, R_2$ and use it to calculate the potential energy and probability density function in terms of center-of-mass (c.m.) height $h_{c.m.}$ and plane angle $\theta_p$. The force $F(R, h)$ on one of the spheres is given by \autoref{eq:model_net}. The net force $F_{DB}(R_1, R_2, h_{c.m.}, \theta_p)$ is then given by
\begin{align}
     F_{DB} &= F(R_1, h_1) + F(R_2, h_2),\label{eq:dbf_uneq} \\
     h_1 &= h_{c.m.} + \frac{R_2^3 (R_1+R_2) \sin{\theta_p}}{R_1^3+R_2^3} \label{eq:h1} \\
     h_2 &= h_1 - (R_1+R_2) \sin{\theta_p} \label{eq:h2}
\end{align} \autoref{eq:dbf_uneq} can be integrated to give the potential energy $\phi_{DB}(R_1, R_2, h_{c.m.}, \theta_p)$
\begin{align}
    \phi_{DB} &= \phi_{DB,g} + \phi_{DB,e} \label{eq:uneqphi} \\
    \phi_{DB,g} &= -\left(F_{g}(R_1)h_1 + F_{g}(R_2)h_2\right) \label{eq:uneqphig}\\
    \phi_{DB,e} &=  \frac{B(R_1)}{\kappa} \exp\left[-\kappa h_1 \right] + \frac{B(R_2)}{\kappa} \exp\left[-\kappa h_2 \right]. \label{eq:uneqphie}
\end{align}
This potential can be derived with respect to the $h_{c.m.}$ to obtain the force and to $\theta_p$ to obtain the torque. We subsequently used the potential to derive the appropriate height distribution for the dumbbell c.m. $p_{DB}(R_1, R_2, h_{c.m.}, \theta_p)$ up to a normalization constant,
\begin{align}
    &p_{DB}(R_1, R_2, h_{c.m.}, \theta_p) \propto K\exp\left[-\frac{\phi_{DB}}{k_B T}\right]. \label{eq:uneqp}
\end{align}
$K$ represents the particle-wall hard-core interaction potential contribution to the Boltzmann weight: $K=1$ if both spheres of the dumbbell are above the wall; otherwise $K=0$.

We show the results of \Crefrange{eq:uneqphig}{eq:uneqp} in \autoref{fig:appendix1} and \autoref{fig:appendix2} for dumbbells of the same volume as the \SI{2.2}{\um} and \SI{4.2}{\um} dumbbells, respectively. The individual contributions of the gravitational and electrostatic potential to the net potential energy are shown in the first and second column, respectively, in \autoref{fig:appendix1} (dumbbells of the same volume as the \SI{2.2}{\um} dumbbells) and \autoref{fig:appendix2} (dumbbells of the same volume as the \SI{4.2}{\um} dumbbells). It is clear that the electrostatic potential is not negligible compared to the gravitational potential, therefore, the height from the surface is greatly influenced by electrostatic repulsion despite the relatively short Debye length (on the order of \SI{150}{\nm}). We have calculated two experimentally relevant size dispersities: an experimentally relevant \SI{5}{\percent} size dispersity and a highly anisotropic dumbbell (snowman particle) for which $R_2\approx 2R_1$. We have chosen the $R_1, R_2$ in such a way that the total mass of the dumbbell is the same as the $R_1=R_2=R$ case. As a convention, positive angles denote the orientation where the sphere of the smaller radius $R_1$ is higher than the sphere of the larger radius $R_2$, as given in \Crefrange{eq:h1}{eq:h2}.

Compared to the case where both spheres are equal, increasing the size dispersity between the two spheres has two effects: firstly, the distribution around $\theta_p = 0$ is no longer symmetric, as shown in \autoref{fig:appendix1} and \autoref{fig:appendix2}d-k. Secondly, a larger range of both angles and c.m. heights become accessible.

\bibliography{references.bib}
\bibliographystyle{unsrtnat}

\end{document}